\documentclass[fleqn,10pt]{wlscirep}
\usepackage[utf8]{inputenc}
\usepackage[T1]{fontenc}
\usepackage{comment}

\usepackage{enumitem}
\usepackage{amsmath}
\usepackage{amssymb}
\usepackage{bm}
\usepackage{bbm}
\usepackage{multirow}
\usepackage{esvect}
\usepackage{makecell}
\DeclareMathOperator*{\Min}{\text{\textbf{minimize}}}
\DeclareMathOperator*{\Max}{\text{\textbf{maximize}}}
\DeclareMathAlphabet{\mathcal}{OMS}{cmsy}{m}{n}

\DeclareMathOperator*{\Argmax}{\text{argmax}}

\renewcommand{\thesubsection}{M.\arabic{subsection}}
\newcommand{\Methodsubsection}[1]{%
  \refstepcounter{subsection}%
  \subsection*{M.\arabic{subsection} #1}%
}

\usepackage[symbol]{footmisc}

\usepackage{csquotes}

\title{Advancing network resilience theories with symbolized reinforcement learning}

\author[1,2]{Yu Zheng}
\author[1,2,*]{Jingtao Ding}
\author[1,2]{Depeng Jin}
\author[3,4]{Jianxi Gao}
\author[1,2,*]{Yong Li}
\affil[1]{Beijing National Research Center for Information Science and Technology, Tsinghua University, Beijing, China.}
\affil[2]{Department of Electronic Engineering, Tsinghua University, Beijing, China.}
\affil[3]{Department of Computer Science, Rensselaer Polytechnic Institute, Troy, NY, USA.}
\affil[4]{Network Science and Technology Center, Rensselaer Polytechnic Institute, Troy, NY, USA.}

\affil[*]{corresponding author: dingjt15@tsinghua.org.cn; liyong07@tsinghua.edu.cn}

\begin{abstract}

Many complex networks display remarkable resilience under external perturbations, internal failures and environmental changes, yet they can swiftly deteriorate into dysfunction upon the removal of a few keystone nodes. 
Discovering theories that measure network resilience offers the potential to prevent catastrophic collapses---from species extinctions to financial crises---with profound implications for real-world systems.
Current resilience theories address the problem from a single perspective of topology, neglecting the crucial role of system dynamics, due to the intrinsic complexity of the coupling between topology and dynamics which exceeds the capabilities of human analytical methods.
Here, we report an automatic method for resilience theory discovery, which learns from how AI solves a complicated network dismantling problem and symbolizes its network attack strategies into theoretical formulas.
This proposed self-inductive approach discovers the first resilience theory that accounts for both topology and dynamics, highlighting how the correlation between node degree and state shapes overall network resilience, and offering insights for designing early warning signals of systematic collapses.
Additionally, our approach discovers formulas that refine existing well-established resilience theories with over 37.5\% improvement in accuracy, significantly advancing human understanding of complex networks with AI.

\end{abstract}
\begin{document}

\flushbottom
\maketitle

Resilience, an essential property of complex networks, describes their ability to maintain basic functionality when internal failures and external disturbances occur\cite{buldyrev2010catastrophic,gao2016universal,albert2000error}.
The loss of resilience can lead to catastrophic collapses, often caused by the removal of a few keystone nodes vital to the system's functionality\cite{albert2000error,fan2020finding,sanhedrai2022reviving}. 
For example, the malfunction of a few hub nodes can fragment a communication network, disabling its global information-carrying capacity\cite{albert2000error}. 
Similarly, knocking out a few genes might suppress the expression of the entire cellular network, leading to cell death\cite{gao2016universal}. 
Given the pervasive presence of complex networks and the frequent occurrence of perturbations that compromise resilience in real-world systems, it is crucial to understand how network structure and the underlying dynamics influence their overall system resilience\cite{artime2024robustness}.

In a long-standing effort, physicists have developed a series of network resilience theories to identify keystone nodes whose removal triggers the fastest loss of resilience, aiming to rigorously assess network resilience\cite{albert2000error,gao2016universal,artime2024robustness}.
These approaches\cite{artime2024robustness}---ranging from manually derived formulas including degree centrality\cite{albert2000error} (DC: $d_i$), collective influence\cite{morone2015influence} (CI: $(d_i-1)\sum_{j \in \partial Ball(i, l)}{(d_j - 1)}$), and resilience centrality\cite{gao2016universal,zhang2020resilience} (RC: $2\bar{d_i}+d_i(d_i-2\beta)$) to recently emerged data-driven machine learning methods\cite{grassia2021machine,fan2020finding}---predominantly focus on the topological structure of networks, while leaving a significant gap in addressing the dynamics that drive system functionality\footnote[1]{Here $d_i$ and $\bar{d_i}$ denote the degree and the average neighbor degree of node $i$. $Ball(i, l)$ is the set of nodes inside a ball of radius $l$ (defined as the shortest path) around node $i$, and $\partial Ball(i, l)$ is the frontier of the ball. $\beta=\langle d \rangle + \frac{\sigma^2}{\langle d \rangle}$ where $\langle d \rangle$ and $\sigma$ are expectation and variance of the degree distribution.}.
Due to the complex coupling of network topology and dynamics, advancing resilience theories is extremely challenging, exceeding the limits of current analytical approaches that often oversimplify or overlook network dynamics.
Moreover, the vast solution space involved in identifying keystone nodes from large networks and deriving mathematical formulas requires efficient search capabilities, rendering current approaches intractable.

AI, emerging as an indispensable tool for addressing challenges beyond human capability, holds vast untapped potential to drive theoretical advancements by solving complex problems.
In this work, we explore the possibility in advancing network resilience theories.
Specifically, we use reinforcement learning to automatically reveal insightful resilience indicators and derive principled topological and dynamical characteristics that identify keystone nodes.
Here, resilience theory discovery is embedded into a network dismantling process where the objective is to induce a network collapse by removing the minimal number of nodes.
The designed symbolized reinforcement learning approach not only achieves optimal network dismantling but also learns from how it solves this complicated problem, making theoretical advancement in a self-inductive manner for the first time (see Figure \ref{fig:framework}a).
To address the enormous solution space, we develop a deep reinforcement learning (DRL) agent that utilizes neural networks to guide the search of keystone nodes as attack targets. 
The agent is then symbolized into mathematical formulas revealing its node selection strategy.
The discovered resilience theories, established based on these symbolized formulas, can be directly applied to identify empirically optimal keystone nodes and, more importantly, to characterize and quantify network resilience.
Remarkably, unlike previous theories\cite{albert2000error,gao2016universal,zhang2020resilience} that focus solely on topological resilience, our approach breaks new ground in network science by discovering the first resilience theory that seamlessly integrates both topology and dynamics, encapsulated in an elegant formula $d \cdot s$ representing the product of degree and state.
This dynamical resilience theory provides a concise yet precise measure of individual nodes' contributions to system resilience, surpassing current topological resilience theories by over 74.1\% in accurately identifying keystone nodes.

Our discovered theory also leads to the first network-level resilience indicator, the correlation between degree and state distributions, $\texttt{corr}(d, s)$, which serves as a potential cornerstone of resilience in many complex networks, even for networks previously considered vulnerable to attacks\cite{albert2000error}.
The effectiveness of this resilience indicator is empirically verified in real-world ecology and supply chain networks, offering a novel perspective that explains why natural systems are more resilient to perturbations than man-made ones\cite{ulgiati1998monitoring}.
Whereas existing data-driven machine learning approaches output keystone nodes directly with black-box neural networks, our approach instead sets up the first paradigm that discovers the underlying resilience theories, marking a notable milestone in theoretical advancement using AI.

\section*{Symbolized reinforcement learning: a self-inductive approach}
\subsection*{Identifying keystone nodes with DRL agent}
Identifying keystone nodes that determine the resilience of a network is challenging both theoretically and computationally.
Theoretically, the resilience function resides on manifolds within a high-dimensional space defined by the complex coupling of network topology and non-linear, heterogeneous system dynamics, which is often non-analytic.
Computationally, selecting a subset of nodes from a large network is well-known to be an NP-hard problem\cite{lalou2018critical}, with an enormous solution space of size $\frac{N!}{(N-K)!}$, which grows super-exponentially as the network size $N$ and the number of selected nodes $K$ increase.
For instance, choosing 50 nodes from a 500-node network results in over $10^{133}$ possible combinations, rendering exhaustive search impractical.

We propose to identify keystone nodes recursively with a DRL agent, which removes one node at each step until the network loses resilience (see Figure \ref{fig:framework}b and Section \ref{method::overview}-\ref{method::search}).
Specifically, we design a graph neural network (GNN) tailored to encode rich features of both network topology and complex dynamics into node representations, effectively capturing the high-dimensional, non-linear, and heterogeneous resilience function.
Informed by the learned node representations, the agent calculates the probability of selecting different nodes and estimates the system resilience with two multi-layer neural networks, \textit{i.e.} policy network and value network (Figure \ref{fig:framework}b).
This approach enables the agent to strategically select nodes at each step, efficiently exploring the enormous solution space by prioritizing high-quality actions.
Therefore, we effectively address the NP-hardness of the problem by searching the minimal keystone nodes whose removal brings the fastest loss of network resilience, rather than exhaustively enumerating all possible solutions.
We train the agent with millions of node removal samples across various networks, and eventually, the agent becomes an expert in identifying keystone nodes for the resilience.
The agent outperforms both current theories\cite{albert2000error,gao2016universal,morone2015influence,zhang2020resilience} and machine learning based approaches\cite{grassia2021machine,fan2020finding} thanks to its efficient search of the solution space and its comprehensive integration of both network topology and dynamics.

\subsection*{Symbolizing DRL agent to discover new network resilience theories}
Network theories are often expressed by mathematical formulas that combine physical variables with operators, but the DRL agent functions as a black-box neural network containing thousands of nearly random parameters without any physical meanings, which is difficult to understand by humans. 
Towards this end, we symbolize the DRL agent to reveal its internal rules of identifying keystone nodes, deciphering the mysterious black box into human-comprehensible mathematical formulas.

The rich input topological and dynamical features and feasible mathematical operators result in an infinite number of possible formulas that could explain the DRL agent, rendering theory discovery extremely challenging.
To narrow down the search space, we distill the DRL agent to focus only on its key ingredients by attributing its prediction to individual input features (see Figure \ref{fig:framework}c and Section \ref{method::distill}).
We achieve this by learning a differentiable feature selection mask that maximizes the mutual information between the selected features and the model's prediction while simultaneously minimizing the number of included features.
This process results in a sparse set of important features that play dominant roles in the agent's prediction.
The distillation effectively simplifies the DRL agent and allows for theory discovery within a streamlined subspace of mathematical formulas, considering only the most informative features.

To generate human-understandable and physically insightful formulas, we develop an evolutionary algorithm to explore the enormous mathematical expression space, mapping the important features to identified keystone nodes towards higher regression accuracy and lower expression complexity, rather than perfectly fitting the data which can lead to over-complicated formulas (see Figure \ref{fig:framework}d and Section \ref{method::discover}-\ref{method::refine}).
We design an iterative refinement strategy that repeats the evolutionary process with a decreasing set of operators, progressively narrowing the search space based on obtained results.
Ultimately, this process discovers the underlying rules of how the DRL agent assesses the importance of individual nodes, transforming the black-box model into candidate mathematical formulas.
We involve human experts to pick the most reasonable formulas from the top performed candidates\cite{davies2021advancing}, where reasonable refers to formulas that possess physical significance or can be explained by established theories\cite{gao2016universal}.

\vspace{15px}
\noindent Conceptually, as illustrated in Figure \ref{fig:framework}a, our approach introduces a new self-inductive paradigm---first solve the complex network dismantling problem using DRL, and then learn from how DRL solves the problem by symbolizing its node selection strategies into resilience theories. 
The established theories can be directly applied to identify keystone nodes and predict network resilience.
Now, we demonstrate how this framework has successfully achieved theoretical advancements facing real high-dimensional and non-linear systems that were previously considered intractable with current analytic tools.

\section*{Results}

\subsection*{Resilience theory accounting for both topology and dynamics}

Existing theories\cite{albert2000error,morone2015influence,clusella2016immunization,gao2016universal,sanhedrai2022reviving,zhang2020resilience} only reveal the relationship between resilience and network topology, disregarding system dynamics.
However, the interplay between topology and dynamics is pervasive in real-world systems, yet its complexity exceeds beyond the capacity of current analytical tools\cite{gao2016universal}, leaving resilience theories that account for both aspects still unknown.
We begin with a typical example of widely investigated gene regulatory networks driven by Michaelis–Menten dynamics\cite{karlebach2008modelling,sanhedrai2022reviving,gao2016universal} which exhibit complex non-linearity and heterogeneity, \textit{i.e.} different nodes have different dynamical parameters\cite{barzel2013universality,barzel2015constructing,boccaletti2006complex,karlebach2008modelling} (see Section \ref{method::cellular}).
These networks function adaptively by adjusting back to a desired steady state despite external perturbations, while they could lose resilience after the removal of keystone nodes which irreversibly pushes the node states into dysfunction that deviates from the desired state\cite{gao2016universal,sanhedrai2022reviving,laurence2019spectral,wu2023rigorous}.

Using our framework, the DRL agent strategically removes nodes based on features of both topology and dynamics until the network collapses, \textit{i.e.} node states approaching zero, which indicates cell death.
We then symbolize the DRL agent, eventually discovering a new theoretical formula assessing the contribution of individual nodes to network resilience, $d \cdot s$, representing the product of a node's degree and steady state\footnote[1]{Here $d$ is weighted degree representing the most general case that applies to both weighted and unweighted networks.}.
To validate the universality of the discovered theory in different complex networks, we test it on two distinct systems, including gene regulatory networks driven by the Michaelis–Menten dynamics and neuronal networks\cite{barzel2013universality,barzel2015constructing,sanhedrai2022reviving} driven by the Wilson–Cowan dynamics\cite{wilson1972excitatory,wilson1973mathematical} (see Section \ref{method::neuronal}).
As the first resilience theory accounting for both network topology and dynamics, the discovered $d \cdot s$ formula accurately identifies keystone nodes in real-world networks across various dynamics, topologies, and sizes, as illustrated in Table \ref{tab:result}.
In contrast, existing theoretical approaches (DC\cite{albert2000error}, RC\cite{zhang2020resilience}, CI\cite{morone2015influence}, GND\cite{ren2019generalized}, EI\cite{clusella2016immunization}, and CoreHD\cite{zdeborova2016fast}) and AI models (GDM\cite{grassia2021machine} and FINDER\cite{fan2020finding}) fail to capture network dynamics, leading to 56.0\% more node removal on average---with a maximum of 74.1\%---for gene regulatory networks, and 19.9\% on average for the different neuronal networks.
Full results are provided in Supplementary Note 1-2, Supplementary Table 1-2, and Supplementary Figure 1-3, where the discovered theory constantly offers the most accurate identification of keystone nodes with up to 77.8\% relative improvements.

To comprehensively evaluate the generalizability and robustness of the discovered resilience theory across different types of network structures, we leverage well-established network models\cite{boccaletti2006complex}, including erd\H{o}s-r\'enyi (ER) networks, Barab\'asi--Albert (BA) networks, random partition (RP) networks with communities, and Watts–Strogatz small-world (SW) networks.
Across these network models, the $d \cdot s$ theory demonstrates universal effectiveness by consistently providing precise identification of keystone nodes.
It substantially reduces the number of removed nodes by over 43.3\% for cellular networks, and achieves an even greater improvements of over 56.2\% for neuronal networks (see Figure \ref{fig:result1}a-d, full results in Supplementary Note 3-4 and Supplementary Figure 4-6).
Moreover, we compare the identified keystone nodes with ground-truth values obtained by exhaustive search, illustrating the error relative to the optimal value for different methods in Figure \ref{fig:result1}e-f.
Specifically, existing methods (DC\cite{albert2000error}, RC\cite{zhang2020resilience}, and FINDER\cite{fan2020finding}) exhibit significant errors, particularly as network size increases, with the largest overestimation of keystone nodes by 7.5\%.
In contrast, the $d \cdot s$ formula constantly achieves near-zero errors in removal cost regardless of network size, highlighting the optimality of the discovered resilience theory.

The $d \cdot s$ formula reveals that network resilience is characterized by two macroscopic characteristics, degree $d$ and state $s$, which capture a node's influence on system resilience from the perspective of network topology and dynamics, respectively.
A higher $d$ implies a broader scope of influence, suggesting that the node is more densely connected within the network, while a higher $s$ signifies a stronger and more lasting impact the node has on the entire system.
To our knowledge, the discovered $d \cdot s$ formula is the first resilience theory that accounts for the coupling of network topology and dynamics, filling an important gap in this field to expand our understanding of network resilience and marking an unprecedented milestone beyond existing theories and AI-driven approaches.

\subsection*{Towards network-level resilience via $d \cdot s$}

In practical applications, a systematic theory that captures resilience at the entire network level is still unknown yet highly desired, as it addresses a critical question: which networks are more resilient?
This prompts us to extend the discovered node-level $d \cdot s$ formula to a network-level theory.
Essentially, $d \cdot s$ indicates that hub nodes with high $d$ may not be dominant in maintaining a network's functionality, unless they also exhibit high states $s$.
Therefore, the correlation between degree and state distributions significantly influences network resilience.
Towards this end, we investigate network-level resilience---measured by the number of nodes removed for collapse---under varying correlation of degree and state distributions, using the widely studied scale-free BA networks and homogeneous ER networks.

As predicted in Figure \ref{fig:result2}a-b, networks are least resilient when degree and state distributions are positively correlated (measured by high Pearson correlation between the distribution of node degree $d$ and node state $s$, \textit{i.e.}, $\texttt{corr}(d, s)$), where hub nodes become dominant players again, rendering the system fragile once these hubs are removed.
Surprisingly, as the correlation decreases (lower $\texttt{corr}(d, s)$), networks become substantially more resilient.
For instance in Figure \ref{fig:result2}a-b, the number of removed nodes increases by over 183\% and 300\% when decorrelating degree and state distributions.
The surprising relation between degree-state correlation and network resilience can be explained by the distribution of $d \cdot s$.
Specifically, under conditions where degree $d$ is decorrelated with $s$, the risk of network collapse, indicated by the value of $d \cdot s$, is not concentrated on hub nodes but rather distributed more uniformly across a wide range of nodes, thereby requiring more nodes to be removed to induce the loss of resilience (Figure \ref{fig:result2}c-d).
While classical resilience theories\cite{albert2000error,gao2016universal} address resilience using only topological characteristics, we find that the correlation between degree and state distributions---which considers both topology and dynamics---can effectively predict network-level resilience, offering a new perspective to understand real-world complex systems.

To further validate the predictive ability of the derived network-level resilience proxy in practice, we study two real-world networks, supply chain networks and ecology networks, where the state $s$ represents throughput of companies and specie abundance, respectively.
Notably, we observe that ecology networks generally display lower correlation between degree and state distributions compared to supply chain networks, with the average difference in $\texttt{corr}(d, s)$ exceeding 55.9\% (see Figure \ref{fig:result2}e).
This substantial gap in degree-state correlation explains why natural systems tend to be more resilient than artificial ones\cite{ulgiati1998monitoring}.
Moreover, we observe that in supply chain networks, the degree-state correlation decreases with network size, with a substantial drop in $\texttt{corr}(d, s)$ by 38.8\% (see Figure \ref{fig:result2}f). This suggests that incorporating diverse additional suppliers enhances the resilience of supply chains\cite{tukamuhabwa2015supply}.
These findings from real-world supply chain and ecology networks further confirm that the correlation between degree and state distribution ($\texttt{corr}(d, s)$) indeed serves as a robust and practical indicator for evaluating network-level resilience.

\subsection*{Early warning based on the discovered resilience theory}
Predicting the impact of future failures and attacks is crucial since the loss of resilience is frequently observed in many systems\cite{dakos2014critical,grassia2021machine}.
Early warning is one such application, designed to monitor a network's proximity to critical tipping points with a signal—--typically ranging between 0 and 1—--that quantifies the tolerable damage the network suffers before losing its resilience.
Based on the discovered resilience theory, we design a new early warning signal, 
\begin{align}
    \alpha = \frac{\sum_{i \in V}d_i \cdot s_i}{\sum_{i \in V_c}d_i \cdot s_i}
\end{align}
where $V$ and $V_c$ denote the currently lost nodes and the identified keystone nodes whose removal renders the system non-resilient.
Therefore, $\alpha$ will quickly approach 1 when nodes playing critical roles disconnect, suggesting an abrupt disruption.

The early warning signal $\alpha$ enables precise tracking of resilience in vulnerable real-world systems and provides critical insights on the optimal timing for interventions to prevent further resilience loss.
We employ $\alpha$ for early warning under attack using the most commonly adopted DC\cite{albert2000error} approach where hubs are damaged first, as well as the discovered $d \cdot s$ approach.
In Figure \ref{fig:result3}, we illustrate network functionality and early warning signal under both attack strategies.
Surprisingly, although the network functionality decreases slowly until a sudden collapse, the early warning signal grows much faster when keystone nodes are attacked, indicating that the network has been heavily damaged.
Particularly, we highlight the first response time\cite{grassia2021machine} at an early warning signal of 50\% ($\alpha$ = 0.5).
In Figure \ref{fig:result3}c, the network undergoes an abrupt drop in functionality, while $\alpha$ reaches 0.5 long before the system is disrupted, with the first response time 38\% ahead of the actual collapse, triggering alarms for potential systematic failures and allowing for timely emergency responses.
More results on applying $d \cdot s$ to enhance network resilience are provided in Supplementary Note 7 and Supplementary Figure 9-10, demonstrating that protecting just 3 nodes with the largest $d \cdot s$ values can increase network resilience by over 224\%.
These findings underscore the practical value of the discovered theory, offering actionable insights to safeguard real-world systems from detrimental disruptions.

\subsection*{Diverse use cases of symbolized reinforcement learning}
Physicists have developed several network resilience theories\cite{albert2000error,cohen2001breakdown,gao2016universal}, yet through extensive manual analysis and derivations based on assumptions that often prove invalid in practical systems.
By leveraging reinforcement learning to interact with real complex networks and symbolizing AI models into mathematical formulas, our self-inductive framework offers a transformative approach to enhancing existing resilience theories without introducing any assumptions or manual derivation.
Here, we show how the framework has automatically discovered refined network resilience theories over established ones.

\subsubsection*{Topological resilience}
As mentioned before, existing network resilience theories largely focus on the topological structural---network functionality sustains as long as the topology remains connected---such as Internet and WWW.
In this case, network resilience is usually measured by the size of the greatest connected component (GCC)\cite{albert2000error,morone2015influence,artime2024robustness}, where a network fully loses its resilience until the GCC size approaches 1 indicating topology disintegration\cite{schneider2011mitigation,fan2020finding}.
Albert, Jeong and Barab\'asi\cite{albert2000error} first investigated structural resilience in 2000, where they attacked networks according to node degree (DC) that prioritizes hubs.
Since then, research on structural resilience has been emerging, evidenced by well-established theoretical formulas\cite{morone2015influence,clusella2016immunization}, computational heuristics\cite{zdeborova2016fast,ren2019generalized,artime2024robustness}, and black-box models\cite{ren2019generalized,fan2020finding,grassia2021machine,engsig2024domirank,artime2024robustness} for identifying keystone nodes of these networks.

We train our DRL agent using widely adopted BA networks, whose scale-free properties mirror the degree distribution found in real-world systems\cite{barabasi1999emergence}. 
At each step, the agent receives a reward of the accumulated normalized connectivity\cite{schneider2011mitigation,fan2020finding}, encouraged to disconnect the network with minimal node removal.
We then symbolize the DRL agent, and discover a novel theoretical formula $\frac{d^2}{\bar{d}}$, where $d$ and $\bar{d}$ represent the degree and the average neighbor degree, respectively.
We evaluate on real-world social networks as illustrated in Table \ref{tab:null-dynamics}.
Specifically, the discovered $\frac{d^2}{\bar{d}}$ formula becomes the new state-of-the-art with surprising accuracy in identifying keystone nodes for structural resilience, outperforming CI\cite{morone2015influence} by 1.72\% on average, with a maximum of 3.53\% improvement, and consistently achieving the least number of node removal.
Particularly, the DRL method FINDER\cite{fan2020finding}, the best-known AI model, performs substantially worse than our discovered $\frac{d^2}{\bar{d}}$ theory by 16.16\% on average.

As existing theories either consider individual nodes\cite{albert2000error} or collective effect\cite{morone2015influence}, our discovered formula significantly extends them by incorporating the competition between neighbors.
Specifically, rewriting $\frac{d^2}{\bar{d}}$ as $\frac{d}{\bar{d}} \cdot d$, the discovered formula suggests that a node's influence on network resilience is determined by not only the degree of itself, but also its competitiveness over its neighbors.
While classical theories such as CI, denoted by $(d_i-1)\sum_{j \in \partial Ball(i, l)}{(d_j - 1)}$, prioritize dense clusters, our discovered theory, in contrast, predicts that hub nodes with both large $d$ and $\bar{d}$ are less critical due to their interchangeability with neighbors, reflected by their lower $\frac{d}{\bar{d}}$.
Instead, nodes with large $d$ but small $\bar{d}$ effectively connect weakly linked components, playing more dominant roles due to higher $\frac{d}{\bar{d}}$.
The discovery of $\frac{d^2}{\bar{d}}$, refining the well-established theories with superior accuracy, offers important and novel insights that push the boundary of human understanding on structural resilience, showcasing the large potential of AI in advancing classical network theories.

\subsubsection*{Improving over resilience centrality}

Gao, Barzel, and Barab\'asi\cite{gao2016universal} studied the functional resilience of complex networks by introducing linear and homogeneous assumptions on networks dynamics, \textit{i.e.} all nodes share the same dynamical parameters.
Through dimension reduction analysis, they obtained a universal $\beta$ theory quantifying network resilience, based on which resilience centrality (RC\cite{zhang2020resilience}: $2\bar{d}+d(d-2\beta)$), was derived to identify keystone nodes.

In this scenario featuring homogeneous system dynamics, we train our DRL agent on gene regulatory dynamics, and evaluate on both real-world gene regulatory and neuronal networks\cite{karlebach2008modelling,barzel2013universality,barzel2015constructing,gao2016universal,sanhedrai2022reviving}.
After symbolizing the agent to decipher its node selection strategy, we discover a new resilience formula, $2\bar{d}+d(\bar{d}-2\beta)$.
As shown in Table \ref{tab:homo-dynamics}, our discovered formula indeed outperforms RC for both network dynamics, achieving an average improvement of 22.5\% and 11.9\% in accuracy of the identified keystone nodes for two types of gene regulatory networks, with a notable maximum improvement of 37.5\%.

Building upon the well-established $\beta$ theory\cite{gao2016universal} and RC theory\cite{zhang2020resilience}, our discovered $2\bar{d}+d(\bar{d}-2\beta)$ formula suggests that the ability of a node to affect network resilience relates to both its degree and nearest-neighbor degree.
Notably, this theory improves upon RC by modifying one term, replacing $\bar{d}-2\beta$ with $d-2\beta$.
This change suggests that a node's degree has a positive impact on its importance to network resilience when $\bar{d}$ exceeds $2\beta$, rather than $d$ surpassing $\beta$, as previously suggested.
In fact, RC introduces approximations during its derivation, which can lead to non-neglectable inaccuracies.
The performance improvements highlight that, in addition to discovering entirely new theories, our framework can make subtle yet significant advancements to existing ones, deepening our understanding on network resilience.

\section*{Discussion}

In this work, we present symbolized reinforcement learning, a new \textit{AI for science} framework which for the first time achieves theoretical advancements in complex networks.
Featuring a self-inductive manner, the key difference between our approach and prevailing AI methods\cite{fan2020finding,grassia2021machine,engsig2024domirank,fawzi2022discovering} lies in the non-trivial transformation from a black-box DRL agent to insightful and human-understandable mathematical formulas.
Notably, our approach is capable of refining classical network theories traditionally derived through extensive derivations, such as the complicated dimension reduction analysis used in the universal $\beta$ formula\cite{gao2016universal} (Supplementary Note 5, Supplementary Table 4-5 and Supplementary Figure 7).
Our results demonstrate that AI holds significant untapped potential for pushing the theoretical boundary of complexity science.

Using our framework, we have successfully established a comprehensive set of theories that expand human knowledge on network resilience.
On the one hand, our framework refines existing resilience theories in analyzable systems, which has attracted substantial research efforts in the past 10 years, including pure topological resilience and functional resilience with homogeneous dynamics.
On the other hand, our framework discovers entirely new theories in systems that were believed to be intractable using current analytical tools, leading to the first resilience theory accounting for the coupling of network topology and heterogeneous dynamics.
The discovered comprehensive set of formulas represent significant advancements in resilience theories which have far-reaching implications for real-world complex systems, enabling the development of early warning signals to prevent catastrophic collapses and offering valuable insights for designing practical systems resilient against perturbations.

The benefits of our proposed symbolized AI framework are multi-fold.
It can help achieve superior generalization towards unseen data, as evidenced by Supplementary Note 6 and Supplementary Figure 8a where the symbolized $d \cdot s$ formula outperforms its DRL agent precursor on networks that are not included in the training data.
Additionally, it offers super-fast inference speed, which sidesteps the need for complete computations of neural networks.
For example in Supplementary Note 6 and Supplementary Figure 8b, the $d \cdot s$ formula reduces inference time by over 34.5\%, with speedup exceeding 50.1\% for large networks.
This fusion of generalizability and efficiency highlights the transformative potential of symbolized AI in both theoretical advancements and practical applications in complex networks.

A promising direction for future research is to explore the application of our approach beyond network resilience, such as discovering novel theories on network growth\cite{albert2002statistical,watts1998collective} and percolation\cite{callaway2000network}.
Moreover, our self-inductive framework has far-reaching implications in the \textit{AI for science} field.
Besides complexity science, the symbolized reinforcement learning approach can also be extended to inspire innovative discoveries in other domains where theoretical advancements have traditionally been achieved by human experts, such as mathematics\cite{romera2023mathematical,AlphaGeometryTrinh2024}, biology\cite{senior2020improved,jumper2021highly} and materials\cite{merchant2023scaling}.
More broadly, our work could serve as a prototype for AI systems that operate in a human-understandable way, assisting in the acquisition of new scientific insights.

\section*{Methods}

\begin{figure}[t]
\centering
\includegraphics[width=0.9\linewidth]{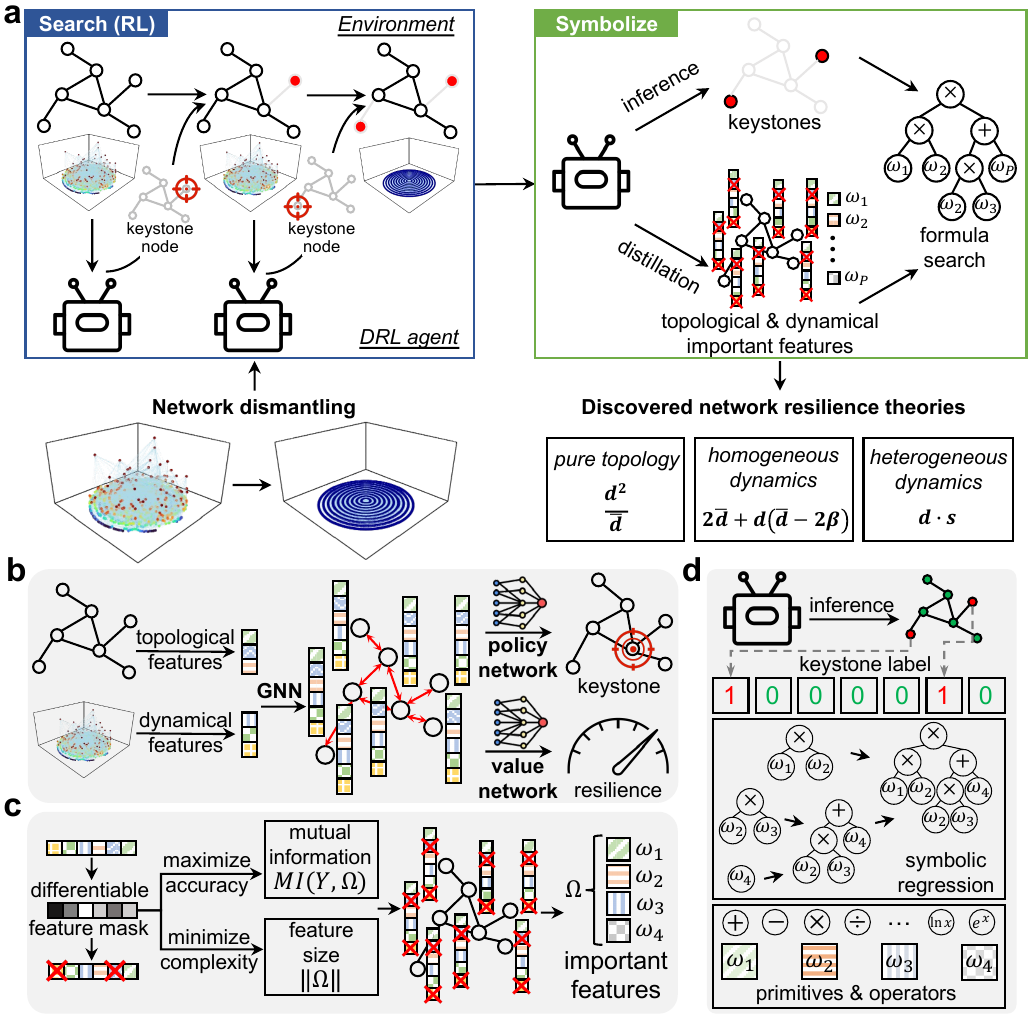}
\caption{
\textbf{a. The proposed self-inductive symbolized reinforcement learning framework.} It first dismantles networks with a DRL agent searching for keystone nodes, then symbolizes the DRL agent with mathematical formulas unraveling its node selection strategy, leading to the discovery of network resilience theories.
\textbf{b. The GNN-based DRL agent.} It encodes rich features of both
network topology and complex dynamics into node representations, based on which it calculates the selection probability of keystone nodes and estimates the network resilience with a policy and value network, respectively.
\textbf{c. Feature distillation.} It attributes the agent's prediction to individual input features with a differentiable feature selection mask maximizing the mutual information between selected features and model prediction as well as minimizing the number of included features.
A sparse set of important features are reserved for further theoretical discovery.
\textbf{d. Theoretical formula generation.} An evolutionary symbolic regression algorithm is developed to map the distilled important features to keystone nodes identified by the DRL agent with algebraic operators, leading to human-understandable mathematical formulas revealing network resilience.
}
\label{fig:framework}
\end{figure}

\Methodsubsection{Overview}\label{method::overview}
In this study, we introduce an \textit{AI for science} framework designed to deepen human understanding in complex networks, particularly in high-dimensional systems characterized by non-linear and heterogeneous dynamics—domains where traditional approaches relying on theoretical analysis often fall short.
The proposed framework is not limited to a specific task but serves as a versatile tool for scientific discovery in complex networks. 
Specifically, we leverage this framework to quantify network resilience and identify associated critical nodes.
As shown in Figure \ref{fig:framework}, our framework employs AI to accomplish the complicated network dismantling task, searching critical nodes with RL, then symbolize itself into new network resilience theories.
In this self-inductive framework, human experts are involved to define the problem and refine the candidate results generated by AI, while leaving the complicated and tedious searching procedure to the AI model\cite{davies2021advancing}.
We now provide a brief overview of resilience theory discovery using our framework (Supplementary Figure 13).
\begin{itemize}[leftmargin=*]
    \item \textbf{Define.} 
    In order to employ AI models, the initial step involves framing the task as a well-defined computational problem with appropriate input and output specifications. 
    Since the ground-truth of keystone nodes can only be obtained by exhaustive search, a practically infeasible endeavor, we approach the task by approximating it as an optimization problem, \textit{i.e.} the network dismantling problem. 
    In specific, we remove nodes from the network to compromise its resilience, with the objective of minimizing the total cost of node removal. 
    The criteria for selecting nodes thus serve as the indicator of node importance.
    \item \textbf{Search.} 
    The formulated network dismantling problem exhibits an enormous solution space, which requires efficient search to obtain a decent strategy.
    Towards this end, we frame the problem as a Markov decision process (MDP), where one node is removed per step until the network loses resilience.
    We design a deep reinforcement learning (RL) agent to solve this MDP, with the object of concluding the episode with the fewest steps.
    We create an environment to simulate high-dimensional networks characterized by non-linear and heterogeneous dynamics.
    This environment evaluates the resilience of the current system and provides feedback to the DRL agent.
    A GNN based state encoder is designed to capture rich topological and dynamical features of nodes, which is shared between a policy network and a value network, with the former selecting nodes for removal and the latter predicting network resilience based on the learned network representations, respectively.
    Through millions of interactions with the environment, the RL agent eventually attains a superhuman level performance in quantifying network resilience and identifying keystone nodes, resulting in a highly accurate solution to the network dismantling problem that is characterized by neural networks.
    \item \textbf{Distill.}
    The DRL model operates as a black box, offering limited explainability regarding how it derives predicted node importance scores, which impedes human comprehension of the contribution of different nodes to the overall network resilience.
    As the node removal decision is made based on node embeddings extracted from rich node features, we distill the DRL model with explainable AI (XAI) techniques to scrutinize feature importance, attributing the model prediction to individual node features.
    Among all input node features, certain features emerge as pivotal that dominate the agent's decision-making process, while others prove insignificant in influencing model predictions.
    Through XAI, we eliminate noise from the complicated DRL model, and distill a streamlined version of the model that retains only the important features.
    \item \textbf{Discover.}
    Ultimately, we aim to decipher the internal mechanisms of the AI model, discovering underlying rules behind critical nodes that are easy to understand by human.
    In pursuit of this goal, we discover mathematical connections between important features and critical nodes with symbolic regression (SR), to derive a physically insightful index of node importance from the DRL agent.
    A diverse set of complex networks is employed to construct a dataset for SR, based on which a mathematical formula is discovered, predicting the nodes selected by the DRL model from the important features distilled by XAI.
    This resultant human understandable physical metric faithfully replicates the behaviors of the DRL model while in the meantime offering superior explainability.
    Through employing SR to correlate the DRL policy with the important node features, we generate valuable insights, further enriching our understanding of what makes the success of AI models.
    \item \textbf{Refine.}
    The above discovery process can generate multiple candidate formulas.
    Meanwhile, the extracted equations may exhibit complexity, harboring redundant terms or unnecessary constant values.
    Therefore, as the concluding step in our proposed \textit{AI for science} framework, human-AI collaboration is employed to refine the generated formulaic results.
    Ultimately, a simplified and concise metric is derived that comprehensively captures the importance of different nodes to the overall network resilience.
    The resulting metric is designed to be easily understood by human, enhancing its utility in practical applications such as network protection and network design.
\end{itemize}
With the outlined framework, we take full advantage of the computational power of AI not only to obtain an accurate model but, more importantly, to generate tangible and insightful formulas about network resilience.
Diverging from prevailing \textit{AI for science} methods that culminate in black-box models, our approach harnesses AI to not only solve the problem but also understand how AI achieves the success, leading to the generation of human-understandable theories. 
Subsequently, we elaborate on the design of our proposed framework.

\Methodsubsection{Define as an optimization problem}\label{method::formulation}
Complex systems can be described as a network of $N$ components (nodes) $V = \{1, \ldots , N\}$ whose activities $\mathbf{x}=(x_1, \ldots, x_N)$ are driven by heterogeneous self-dynamics $\mathbf{F}=\{F_1, \ldots, F_N\}$ and interactions between them $\mathbf{G}=\{\ldots, G_{ij}, \ldots\}$, as characterized by the coupled equations\cite{barzel2013universality,barzel2015constructing}
\begin{align}
    &\frac{\mathrm{d}x_i}{\mathrm{d}t} = F_i(x_i) + \sum_{j=1}^N{A_{ij}G_{ij}(x_i, x_j)},\label{eq:dynamic}
\end{align}
where non-linear dynamic functions\cite{boccaletti2006complex,karlebach2008modelling,sanhedrai2022reviving,meena2023emergent} $F_i$ and $G_{ij}$ are intricately intertwined through the interaction matrix $\mathbf{A}$.
Despite frequent external perturbations and environmental changes that push $\mathbf{x}$ away from its desired functional state $\mathbf{x^H}$, system failures are rarely observed; instead, these systems often display a surprising degree of resilience, being able to adjust back to $\mathbf{x^H}$ and maintain their basic functionality\cite{gao2016universal,sanhedrai2022reviving,laurence2019spectral,wu2023rigorous}.
However, errors on the components (node loss) can lead to an irreversible phase transition, driving the system into a non-resilient basin manifesting dysfunction and breakdowns upon the removal of a subset of nodes $V_c \subset V$, 
\begin{align}
    \text{resilient}\quad &\mathcal{R}(\mathbf{A}, [\mathbf{F}, \mathbf{G}])=1 \quad \Leftrightarrow \quad \mathbf{x}(t\rightarrow \infty) = \mathbf{x^H}, ~~\forall \mathbf{x}(t=0)\\
    \text{non-resilient}\quad &\mathcal{R}(\mathbf{A}[V \setminus V_c], [\mathbf{F}, \mathbf{G}])=0 \quad \Leftrightarrow \quad \mathbf{x}(t\rightarrow \infty) \neq \mathbf{x^H}, ~~\exists \mathbf{x}(t=0)
\end{align}
Here $\mathcal{R}$ measures the resilience of the system which cannot be expressed as a mathematically analytic equation due to the non-linearity and heterogeneity of $[\mathbf{F}, \mathbf{G}]$.
While all nodes influence the functionality of the entire system, the contributions of different nodes to the overall network resilience are not equal due to the prevalent asymmetry property in real-world complex networks. 
Notably, a small subset of critical nodes $V_c \in V$ can have a significant impact on network resilience, and their removal can drastically degrade the network's basic functionality; see Figure \ref{fig:framework}a (bottom-left).
The discovery of critical nodes in network resilience can be formulated as an optimization problem as follows,
\begin{align}
    \Min\limits_{V_c}\quad &\|V_c\|,\label{eq:optimization_start}\\
    \text{\textbf{subject to}}\quad &\mathcal{R}(\mathbf{A}[V \setminus V_c], [\mathbf{F}, \mathbf{G}])=0,\\
    \text{\textbf{where}}\qquad &V_c \subset V \label{eq:optimization_end}
\end{align}
Here, the goal is to minimize the number of removed nodes $\|V_c\|$ to compromise the network's resilience. 
The nodes selected for removal represent the discovered critical nodes $V_c$, and the criteria for selection serves as a useful index for node importance.
The problem (\ref{eq:optimization_start}-\ref{eq:optimization_end}) in its original form requires generating all critical nodes at once which is intractable, thus it is usually transformed to a recursive version as follows,
\begin{align}
    \Min\limits_{\mathcal{Q}}\quad &\kappa,\label{eq:new_optimization_start}\\
    \text{\textbf{subject to}}\quad &\mathcal{R}(\mathbf{A}[V \setminus V_c^\kappa], [\mathbf{F}, \mathbf{G}])=0,\\
    \text{\textbf{where}}\qquad &V_c^0 = \varnothing,\\
    &V_c^\kappa = V_c^{\kappa-1} \cup \{v_c^\kappa\}, \\
    &v_c^\kappa = \Argmax\limits_{i}\{\mathcal{Q}(i; \mathbf{A}[V \setminus V_c^{\kappa-1}], [\mathbf{F}, \mathbf{G}]) \mid i \in V \setminus V_c^{\kappa-1}\}, ~~\kappa=1,2,\ldots \label{eq:new_optimization_end}
\end{align}

\Methodsubsection{Search with DRL}\label{method::search}

\noindent\textbf{MDP definition.}
Finding the optimal set of keystone nodes $V_c$ is a computationally challenging problem \cite{lalou2018critical}, known to be NP-hard\cite{lalou2018critical}, whose ground-truth solutions can only be obtained through exhaustive search, which is infeasible in practical scenarios involving large-scale networks.
To address the challenge brought by the large solution space, we leverage DRL to facilitate efficient search, aiming to train an intelligent agent capable of identifying keystone nodes.
To achieve this, we first reformulate the problem as a Markov Decision Process (MDP), which consists of the following key elements:
\begin{itemize}[leftmargin=*]
    \item \textbf{State.} 
    This describes the current conditions of the complex network, including the topology structure $\mathbf{A}_t=(V_t, E_t)$, and corresponding comprehensive node features considering both topology and dynamics.
    Specifically, we characterize each node with an 11-dimensional feature set, which includes attributes such as \verb|degree|, \verb|maximum edge weight|, \verb|neighbor degree|, \verb|resilience centrality|, \verb|dynamical parameters(3)|, \verb|state|, \verb|state derivative|, \verb|neightbor state|, and \verb|neighbor state derivative|.
    \item \textbf{Action.}
    At each step of the MDP, the DRL agent removes one single node, attempting to fail the network's resilience.
    Therefore the action $a_t$ indicates the selected node for removal at step $t$.
    \item \textbf{Reward.}
    Since the goal is to minimize the number of removed nodes, the reward $r_t$ at each step is set to $-1$, encouraging the process to conclude promptly with the fewest steps.
    The process terminates when the network loses its resilience. 
    The agent is trained to achieve a higher long-term return, which is the cumulative sum of the rewards at each step. 
    Consequently, it learns to identify the most influential nodes for removal, compromising the network's resilience with fewer steps.
    \item \textbf{Transition.}
    It describes how the network changes with the current action, where the node $a_t$ is removed along with all edges connected to it.
    In addition, the removal of the node $a_t$ may make the network disconnected into multiple components.
    In such cases, we focus on the residual greatest connected component (GCC) subsequently in accordance with existing literature \cite{gao2016universal}.
    Therefore, the network $\mathbf{A}_{t}$ transitions to $\mathbf{A}_{t+1}$, which is either $\mathbf{A}_t \setminus a_t$ or $\texttt{GCC}(\mathbf{A}_t \setminus a_t)$, depending on the connectivity of the residual network.
\end{itemize}

Figure \ref{fig:framework}a (top-left) and b and Supplementary Figure 13b demonstrate our proposed DRL approach, where an agent interacts with an environment, with the object of compromising the network's resilience using the fewest steps of node removal.
To begin with, we construct an environment to simulate the complex network.
This environment implements the previously described transition process for each action, thereby altering the network structures.
It generates the state by simulating the trajectories of the network given the current topology $\mathbf{A}_t$ and dynamic functions $[\mathbf{F}, \mathbf{G}]$ described with ordinary derivative equations (ODE) as equations (\ref{eq:dynamic}).
Meanwhile, it evaluates the resilience of the current network by perturbing the states $\mathbf{x}$ and evaluating whether the system can remain at its desired state $\mathbf{x^H}$ or featuring an undesired state $\mathbf{x^L}$, providing rewards to the agent.
Additionally, it terminates the episode once the network loses its resilience.

\noindent\textbf{GNN state encoder.}
We develop an agent consisting of a state encoder for learning node representations, a policy network for selecting nodes for removal, and a value network for estimating network resilience.
To capture the intricate topological and dynamical features, we employ Graph Neural Networks (GNN)\cite{kipf2017semi} to encode the rich node features, resulting in dense embeddings of different nodes.
Specifically, we conduct message passing and neighbor aggregation over the graph, through stacking several linear transformation layers and non-linear activation layers.
The 11-dimensional node features serve as input attributes to the GNN, leading to the generation of initial node embeddings as follows:
\begin{align}
    &\mathbf{X}_i = \eta_1(i) \parallel \eta_2(i) \parallel \cdots \parallel \eta_{11}(i),\label{eq:feature}\\
    &\mathbf{h}_i^0 = \mathrm{tanh}(\mathbf{W}^0 \mathbf{X}_i),
\end{align}
where $\mathbf{X}_i$ denotes the input attributes for node $i$, and $\parallel$ signifies the concatenation of the 11 different features as introduced previously.
We transform the input node attributes with a linear transformation matrix $\mathbf{W}^0$ and the tanh activation function.

We then propagate the node embeddings over the graph and aggregate messages from neighbors to iteratively update node embeddings.
This recursive process involves several graph convolutional layers, as expressed below,
\begin{align}
    \mathbf{h}_i^{l} = \mathrm{tanh}(\mathbf{W}^{l}\sum_{j \in \mathcal{N}(i)\cup\{i\}}{\frac{e_{j,i}}{\sqrt{\hat{d_j}\hat{d_i}}}\mathbf{h}_j^{l-1}}), l=1,2,\cdots,L,
\end{align}
where $\hat{d_i}=1+\sum_{j \in \mathcal{N}(i)}{e_{j,i}}$, and $e_{j,i}$ denotes the edge weight from node $j$ to node $i$.
For each node, this process absorbs information from its neighboring nodes to update its own embedding, utilizing a linear transformation layer $\mathbf{W}^{l}$ and the tanh activation unit.
Through the stacking of multiple graph convolutional layers, we obtain the final node embedding $\mathbf{h}_i^{L}$, where $L$ is a hyper-parameter in our model.
The learned node representations encapsulate both the topological and dynamical features of each node, and meanwhile consider the influence of high-order neighbor nodes.
These semantic embeddings inform subsequent action making and return prediction, with the GNN stated encoder shared between policy and value networks.

\noindent\textbf{Policy and value network.}
We employ the actor-critic approach, with a policy network taking actions and a value network estimating returns at each step.
Specifically, the policy network assigns scores to each node, based on which one single node is selected for removal.
Node scores are computed from the node embeddings, extracted by the GNN state encoder.
On the other hand, the value network takes graph-level representations and predicts the return, which captures the resilience conditions of the current complex system.
To achieve a comprehensive graph-level representation, summarizing the information of the entire network, we use the average of all node embeddings.
For both the policy and value networks, we employ multi-layer perceptrons (MLP) models, expressed as follows,
\begin{align}
    &s_i = \mathrm{MLP}_{p}(\mathbf{h}_i^{L}),\label{eq:policy_network}\\
    &\hat{r_t} = \mathrm{MLP}_{v}(\frac{1}{\lvert V \rvert}\sum_{i \in V}{\mathbf{h}_i^{L}}),\label{eq:value_network}
\end{align}
where $s_i$ is the predicted score for node $i$, $\hat{r_t}$ is the estimated return value, and $\mathrm{MLP}_{p}$ and $\mathrm{MLP}_{v}$ consist of several consecutive linear feed-forward layers and non-linear activation layers.
During training, nodes are selected by sampling based on their corresponding scores, with the selection probability proportional to their scores as follow,
\begin{align}
    \mathrm{Prob}(i) = \frac{e^{s_i}}{\sum_{j \in V}{e^{s_j}}},
\end{align}
where we utilize this softmax function to construct a proper probability distribution.
When applying a well-trained model to identify keystone nodes, we select nodes greedily at each step as follow,
\begin{align}
    a_i = \mathrm{argmax}\{s_i \mid i \in V\},
\end{align}
where the node with the largest predicted score is selected.
It is noteworthy that the score $s_i$ reflects the importance of node $i$, and the estimated value $\hat{r_t}$ indicates how resilient the current network is.
In essence, the policy network, along with the GNN state encoder, functions as a black-box estimator, assessing the contribution of different nodes to the overall network resilience.
Similarly, the value network, coupled with the GNN state encoder, serves as an estimator evaluating the resilience of a complex network.
However, the parameters of the neural networks are challenging to interpret directly by humans.
We will later open up these black-box estimators, unveiling tangible and insightful formulas that are much more perceivable to humans.

\noindent\textbf{Model training.}
We implement both the environment and the agent with the Stable-Baselines3 framework\cite{stable-baselines3}, and train the agent using the Proximal Policy Optimization (PPO) algorithm\cite{schulman2017proximal}.
Specifically, we utilize the following surrogate clipped object as the policy loss, encouraging the policy to make adaptive updates,
\begin{align}
    &r_t(\theta) = \frac{\pi_{\theta_{new}}(a_t \mid s_t)}{\pi_{\theta_{old}}(a_t \mid s_t)},\\
    &\hat{A_t} = Q(s_t, a_t) - v(s_t),\\
    &L_{p}(\theta) = \mathrm{min}(r_t(\theta)\hat{A_t}, \mathrm{clip}(r_t(\theta), 1 - \epsilon, 1 + \epsilon)\hat{A_t}),
\end{align}
where $\theta$ denotes the model parameters, $r_t(\theta)$ is the ratio of the probability under the new policy to the old policy, $\hat{A_t}$ is the estimate advantage at time $t$, and $\epsilon$ is the clip hyper-parameter to prevent large model updates.
With the clipping mechanism, we stabilize the training process while guarantee safe exploration. 
After multiple iterations of policy optimization, the policy gradually approaches optimal solutions with each iteration improving a small step near the original policy.

Due to the enormous solution space, it is impractical for the agent to enumerate all feasible node removal strategies during interactions with the environment.
To address this challenge, we include an entropy loss on the action sampling probability distribution to encourage the policy to be more stochastic as follow,
\begin{align}
    L_{e} = -\mathrm{Entropy}(\{\mathrm{Prob}(i) \mid i \in V\}).
\end{align}
Here entropy encourages the agent to explore a broader range of actions, even those with lower probabilities.
This proves particularly beneficial in the early stages of learning or when facing environments with high uncertainty, where increased exploration aids the agent in discovering better policies and improving its overall performance.
In addition, it helps prevent the policy from becoming excessively deterministic and overfitting to the observed data. 
By penalizing policies that assign very high probabilities to a single action, entropy loss promotes a more balanced distribution over actions.

We adopt mean squared error between the estimated return and the ground-truth to supervise the value network as follow,
\begin{align}
    &R_t = r_{t+1} + \gamma r_{t+2} + \gamma^2 r_{t+3} + \cdots = \sum_{k=0}^{\infty}{\gamma^k r_{t+k+1}},\\
    &L_{v} = \mathrm{MSE}(\hat{r_t}, R_t),
\end{align}
where $R_t$ is the long-term return at time step $t$, a cumulative discounted aggregation of per-step reward, and $\hat{r_t}$ is the predicted return by the value network according to equation (\ref{eq:value_network}).
The agent is optimized jointly by the three loss functions as follow,
\begin{align}
    L = L_{p} + \lambda_1 L_{e} + \lambda_2 L_{v},
\end{align}
where $\lambda_1$ and $\lambda_2$ are hyper-parameters in our model controlling the weight of different loss functions.

We train the DRL model with cellular networks, running the Michaelis–Menten dynamic function on 30 different scale-free topologies, each consisting of 80 nodes.
During training episodes, the environment iterates over the 30 distinct graphs, with the agent interacting by removing one node from the current topology at each step.
The DRL model is trained for 1 million steps, allowing the agent to observe thousands of induced residual topologies after the removal of identified nodes.
Meanwhile, we periodically evaluate the average performance of identifying keystone nodes across the 30 networks to monitor the model training process.
After sufficient training, we obtain an DRL model proficient in searching for keystone nodes of network resilience with superhuman-level accuracy, yet with limited explainability regarding why it selects specific nodes (see Supplementary Note 4.1 and Figure 6a).
Such explainability is crucial for comprehending the success of AI models, ultimately fostering novel insights and enhancing human understanding of network resilience.

\Methodsubsection{Distill with XAI}\label{method::distill}
\noindent\textbf{Binary classifier GNN for distillation.}
The DRL agent is able to accurately identify critical nodes $V_c$ of network resilience, yet it is a black-box model lacking transparency in how it selects these nodes.
Particularly, the node selection decision is made based on learned node embeddings by a complicated GNN model, which involves multiple layers of message propagation and non-linear activation units.
To understand the GNN based policy, we utilize XAI to distill key information from the black-box model.
In specific, we first construct a separate GNN model $\Theta$ by combining the GNN state encoder and the policy network in equations (\ref{eq:feature}-\ref{eq:policy_network}), as expressed below,
\begin{align}
    \hat{y_i} = \mathrm{MLP}_{p}(\mathrm{tanh}(\mathbf{W}^{L}\cdots\mathrm{tanh}(\mathbf{W}^{2}\sum_{j \in \mathcal{N}(i)\cup\{i\}}{\frac{e_{i,i}}{\sqrt{\hat{d_j}\hat{d_i}}}\mathrm{tanh}(\mathbf{W}^{1}\sum_{j \in \mathcal{N}(i)\cup\{i\}}{\frac{e_{i,i}}{\sqrt{\hat{d_j}\hat{d_i}}}\mathrm{tanh}(\mathbf{W}^{0}[\eta_1(i) \parallel \eta_2(i) \parallel \cdots \parallel \eta_{11}(i)]})})))).\label{eq:explain_policy}
\end{align}
Here this separate GNN accomplishes a node classification task, predicting whether each node is a critical node in $V_c$, and $\hat{y_i}$ is the raw predicted score for node $i$.
Stripping the value network and all RL training components, we reserve the major functional module, a GNN based binary classifier, denoted as $\Theta$, that makes predictions for each node.
We set the parameters of this separate GNN to the values of the corresponding layers in the RL agent, which guarantees that it can fully reproduce the results of the RL agent.

\noindent\textbf{Mutual information maximization.}
With this separate GNN binary classifier $\Theta$, we explain how it makes the predictions using a widely acknowledged XAI tool, GNNExplainer\cite{ying2019gnnexplainer}.
In Figure \ref{fig:framework}a (top-right) and c and Supplementary Figure 13c we illustrate the process of distilling the complicated model $\Theta$.
Specifically, we attribute the prediction of GNN to the input node features, investigating the importance of different features.
As introduced previously, we incorporate 11 features regarding network topology and dynamics, among which are important features that are influential to the model prediction and insignificant features that contribute little information to the model.
Therefore, we aim to find a subset of important node features $\Omega=\{\omega_1, \ldots, \omega_P\} \subset \{\eta_1, \ldots, \eta_{11}\}$, such that removing them leads to drastic deterioration of model prediction.
From the perspective of mutual information, it can be expressed as follow,
\begin{align}
    \Max\limits_{\Omega}\quad &MI(Y, \Omega) = H(Y) - H(Y \mid X=\Omega),
\end{align}
where $MI$ calculates the mutual information between the model prediction $Y$ and the important features $\Omega$, and $H$ calculates the entropy.
As the entropy $H(Y)$ for the prediction of the fixed well-trained model $\Theta$ is a constant, we are minimizing the conditional entropy as follow,
\begin{align}
    H(Y \mid X=\Omega) = -\mathbb{E}_{Y \mid X=\Omega}[\mathrm{log}_{P_\Theta}(Y \mid X=\Omega)].
\end{align}
Meanwhile, we would like the identified important features $\Omega$ to be a compact set, containing as few features as possible.

\noindent\textbf{Learning feature selection mask.}
To obtain important features $\Omega$, we learn a feature selection mask $M_s \in \{0, 1\}^{11}$ for each node, through which we calculate the masked feature as $\hat{X_i} = X_i\otimes M_s$.
The binary mask is non-differentiable for back-propagation, thus we smooth it into real values using the sigmoid function: $M_s = \mathrm{sigmoid}(M)$ where $M \in \mathbb{R}^{11}$ and is initialized with an empirical distribution $\mathcal{N}(0, 0.1)$.
With the complete model $\Theta$ and the full input features $X_i$, we can obtain the original model prediction $y$; while using the masked input features $\hat{X_i}$, the model predicts $\hat{y}$ that tends to exhibit errors in comparison to $y$.
Therefore, to identify important features, we optimize $M$ with the following loss functions,
\begin{align}
    &L_\mathrm{MI} = -\sum_{\mathbf{A}}{\sum_{i=1}^{N_\mathbf{A}}{y_i\mathrm{log}\sigma(\hat{y_i}) + (1 - y_i)\mathrm{log}(1 - \sigma(\hat{y_i}))}},\\
    &L_\mathrm{S} = -\sum_{\mathbf{A}}{\sum_{i=1}^{N_\mathbf{A}}{\sum_{j=1}^{11}{M_{ij}\mathrm{log}M_{ij} + (1 - M_{ij})\mathrm{log}(1 - M_{ij})}}},\\
    &L = L_\mathrm{MI} + \lambda_3 L_\mathrm{S},\label{eq:xai_loss}
\end{align}
where we calculate the loss for all $N_\mathbf{A}$ nodes of each training network $\mathbf{A}$.
The first term $L_\mathrm{MI}$ captures whether the masked features can reproduce the model prediction, representing mutual information between model prediction and the selected features.
Meanwhile, the second term $L_\mathrm{S}$ computes the entropy of the feature selection mask, penalizing the inclusion of a large set of important features and encouraging a focus on a sparse subset of features.
We introduce a hyper-parameter $\lambda_3$ to combine these two terms, optimizing the feature selection mask $M$ to achieve high mutual information between the selected features and the model prediction, while simultaneously minimizing the number of included features.

\noindent\textbf{Identifying important features.}
We train the feature selection mask $M$ with the adopted 30 networks $\mathbf{A}$ of 80 nodes for the DRL agent, and finally we obtain a complete mask $M \in \mathbb{R}^{(30 \times 80) \times 11}$ indicating the importance of different features in various networks.
Specifically, if a particular feature $\eta_k$ is important, the corresponding weights in $\mathbf{W}^0$ tend to be large, and masking this feature will lead to drastic performance deterioration.
Therefore, by optimizing the mask $M$ with equation (\ref{eq:xai_loss}), the corresponding multiplication mask score $M_{ijk}$ will be large.
Then we compute the overall feature importance scores by aggregating the results of all 30 networks,
\begin{align}
    S_k = \sum_{i=1}^{30}{\sum_{j=1}^{80}{M_{ijk}}}, \quad i=1,\ldots,11
\end{align}
The final important features $\Omega$ are identified according to the computed importance scores,
\begin{align}
    \Omega = \{\omega_1,\ldots,\omega_P\} = \mathrm{topP}\{S_1, \ldots, S_{11}\}.
\end{align}
With $\Omega$, we distill the intricate GNN model $\Theta$ into a streamlined version, eliminating the noise of it and reserving only the important features that dominate its prediction (see Supplementary Note 4.2 and Figure 6b).

\Methodsubsection{Discover with SR}\label{method::discover}
\noindent\textbf{Symbolic regression mapping $\Omega$ to $V_c$.}
To fully reveal the underlying rules of the DRL model, we discover the hidden mathematical connections between important features $\Omega$ and the identified critical nodes $V_c$, establishing formulaic mapping with symbolic regression (SR) from $\Omega$ to $\Theta(X)$.
In other words, we derive a physically insightful index $\theta$ from $\Theta$ to measure the contribution of different nodes to the network resilience, replacing the complicated GNN model $\Theta$ with a human-understandable formula $\theta$ that achieves comparable or even better performance in solving the problem (\ref{eq:new_optimization_start}-\ref{eq:new_optimization_end}) while displays superior explainability and promises to facilitate downstream applications such as system enhancement and protection.
In Figure \ref{fig:framework}a (top-right) and d and Supplementary Figure 13d we illustrate the equation discovery process given the previously obtained $\Theta$ and $\Omega$.
Specifically, the equation discovery process contains the following key elements,
\begin{itemize}[leftmargin=*]
    \item \textbf{Data.}
    We use the 30 synthetic networks for training the DRL model as the raw data for SR, where the DRL model achieves the best performance in comparison to all existing baselines.
    Specifically, we construct an SR dataset, containing the important node features identified by XAI and labels indicating whether each node is selected by the DRL model.
    \item \textbf{Primitives.}
    They are the base functions for SR, serving as the basic ingredients for constructing mathematical equations that combines these primitives.
    As we have distilled the important features that play pivotal roles in the model prediction---the DRL model also combines these features to measure node importance albeit in a complicated way using neural networks---we adopt these identified important features $\Omega$ as primitives.
    \item \textbf{Target.}
    We would like the derived formula to mimic the behavior of the complicated GNN model $\Theta$, thus we utilize the output logit values for each node as the target for SR.
    Specifically, we use the $\hat{y_i}$ in equation (\ref{eq:explain_policy}) of each node as the target to predict from the node's corresponding important features (primitives) $\Omega$.
    \item \textbf{Operators:} These are the mathematical operations that combine various elements, whether primitives or sub-equations constructed from primitives, to formulate the final mathematical expression.
    We adopt an iterative refinement strategy with a decreasing set of operators.
    Specifically, we start with a rather rich set of common mathematical operators, and progressively narrow down the operator set according to the obtained results.
    The final search involves two operators, addition ($+$) and multiplication ($\times$), which align well with related centrality metrics\cite{zhang2020resilience,artime2024robustness}.
    \item \textbf{Loss.}
    Since the target $\hat{y_i}=\Theta(X_i)$ are logits for selecting critical nodes $V_c$, the absolute values are not as important as the relative order of different nodes.
    In other words, the mapping from $\Omega$ to $\Theta(X)$ is more a binary classification task rather than a regression task.
    Therefore, we employ the Sigmoid loss function,
    \begin{align}
        L_\mathrm{SR} = 1 - \mathrm{tanh}(\sigma(\hat{y_i})\cdot\sigma(\Tilde{y_i})),
    \end{align}
    where $\hat{y_i}$ is the logit predicted by the DRL model $\Theta$ and $\Tilde{y_i}$ is the prediction by the mathematical formula from SR.
\end{itemize}

\noindent\textbf{Evolutionary algorithm.}
Searching for empirical equations combining primitives ($\Omega=\{\omega_1,\ldots,\omega_P\})$ with mathematical operators ($+$, $\times$) is also an NP-hard problem\cite{virgolin2022symbolic}, thus we utilize evolutionary algorithm to efficiently explore the enormous mathematical expression space with the PySR library\cite{cranmer2023interpretable}.
Specifically, we keep 60 populations, each of which contains 40 mathematical equations, which evolve towards higher fitting accuracy and lower expression complexity.
The evolution contains the following key procedures,
\begin{itemize}[leftmargin=*]
    \item \textbf{Tournament.}
    It compares a population of mathematical equations, and generally selects the fittest one as the winner of the tournament with large probability.
    Meanwhile, sub-optimal equations may also be selected through random sampling.
    Specifically, the probability of selecting the current fittest one $p=0.9$, and the whole probability sequence over all equations ranked by their fitness is roughly $p$, $p(1-p)$, $p(1-p)^2$, ...
    The winners of the tournament are chosen for breeding by the following mutation or crossover procedure.
    \item \textbf{Mutation.}
    New individuals are generated by mutation which alters the mathematical equations a little bit to search its neighborhood in the mathematical expression space.
    For instance, we can manipulate one operator or primitive,
    \begin{align}
        &\mathrm{mutation~1:} \quad \omega_1 \times \omega_2 + \omega_3 \quad\rightarrow\quad \omega_1 \times \omega_2 \times \omega_3\\
        &\mathrm{mutation~2:} \quad \omega_1 \times \omega_2 \times \omega_3 \quad\rightarrow\quad \omega_1 \times \omega_2 \times \omega_4\\
        &\mathrm{mutation~3:} \quad \omega_2 \times \omega_3 + \omega_1 \times \omega_4 \quad\rightarrow\quad \omega_1 \times \omega_4
    \end{align}
    Other mutations include changing constant terms, generating an entire new equation, and simplifying the equation.
    \item \textbf{Crossover.}
    Besides unary mutations that occur on one single equation, binary crossovers are performed to mix two winners of the tournament for breeding new equations.
    Specifically, two formulas exchange parts of each other to produce new equations, as the following example,
    \begin{align}
        &\mathrm{crossover:} \quad \omega_1 \times \omega_2 + \omega_3, \quad \omega_2 \times \omega_3 \times \omega_4 \quad\rightarrow\quad \omega_1 \times \omega_2 \times \omega_3 \times \omega_4, \quad \omega_2 + \omega_3
    \end{align}
\end{itemize}
After the evolution consisting of tournament, mutation, and crossover, each formula in the population is simplified and its constant coefficients are optimized.
We run 100 iterations of the evolution loop, during the evolution we keep track of a \textit{hall of fame} $H$, which records the best fitted equation of different expression complexity so far.
The final derived index $\theta$ is chosen from $H$ through trading off accuracy and complexity.
With the derived $\theta$, we achieve comparably or even improved precise identification of keystone nodes $V_c$, while ensuring superior explainability and universality, as the mathematical formula is much more tangible and human-understandable.

\Methodsubsection{Refine the AI generated equations}\label{method::refine}

A list of candidate equations $H$ is obtained (refer to Supplementary Table 3), which calculates the selection probability of a node based on its corresponding features, allowing us to quantify $\kappa$ and $V_c$ in a much more explainable manner compared to the original black-box DRL model $\Theta$.
The equations generated by SR may appear complex in their raw form, hence we refine these formulas based on dimension and simplify them by removing constant terms, as they do not impact the order of different nodes.
Finally, as illustrated in Supplementary Note 4.3 and Supplementary Figure 6c, striking a balance between accuracy and simplicity, we obtain a concise and elegant formula denoted as $d \cdot s$, the product of \verb|degree| and \verb|state|, which resides at the core of the black-box RL agent, offering valuable insights into the individual contributions of each node to the network's resilience.

\Methodsubsection{Cellular dynamics}\label{method::cellular}

We first investigate the cellular dynamics describing regulatory interactions among genes.
The activity of genes are characterized by the following coupled Michaelis-Menten function\cite{karlebach2008modelling,sanhedrai2022reviving,gao2016universal},
\begin{equation}
    \frac{\mathrm{d}x_i}{\mathrm{d}t} = -b_i x_i^f + \sum_{j=1}^N{A_{ij}\frac{x_j^h}{1+x_j^h}},\label{eq:method_regulatory}
\end{equation}
where self-dynamics $F_i = -b_i x_i^f$ and interactions $G_{ij} = \frac{x_j^h}{1+x_j^h}$ exhibit strong heterogeneity and non-linearity.
Specifically, we achieve the dynamical heterogeneity via imposing a power-law to the decay rate $b_i$,
\begin{align}
    f(b, a) = a\cdot b^{a-1},\label{eq:power_law}
\end{align}
where $a$ is a hyper-parameter to control the level of heterogeneity.
We set Hill coefficient $h=2$ for the cooperation level and $f=1$ for degradation accordingly\cite{gao2016universal}.

To measure network resilience, we first compute the steady states of genes according to equation (\ref{eq:method_regulatory}),
\begin{align}
    &\mathbf{x}_{0} = (x_1, \ldots, x_N)_{t=0} = (10, \ldots, 10)\\
    &\mathbf{x}_{T} = \mathbf{x}_{t=0} + \int_0^T -\mathbf{b}\mathbf{x}^f + \mathbf{A}\frac{\mathbf{x}^h}{1 + \mathbf{x}^h}\mathrm{d}\mathbf{x},\label{eq:gene_integration}
\end{align}
where $\mathbf{b}=(b_1, \ldots, b_N)$ is the heterogeneous dynamic parameters and $T=400$ to obtain the steady state.
Following related literature\cite{gao2016universal,sanhedrai2022reviving}, a resilient cellular network indicates positive gene activity while a non-resilient cellular network refers to cell death with no gene activity.
Therefore, the resilience function is defined by the following equation,
\begin{align}
    \mathcal{R}(\mathbf{A}, [\mathbf{F}, \mathbf{G}]) = \begin{cases} 1 & \text{if } \langle \mathbf{x}_T\rangle > 0 \\ 0 & \text{if } \langle \mathbf{x}_T\rangle = 0 \end{cases}\label{eq:resilience_gene}
\end{align}
The resilience function (\ref{eq:resilience_gene}) cannot be expressed as explicit mathematical expression, but can only be calculated through integrating the network dynamics as equation (\ref{eq:gene_integration}), which makes the problem (\ref{eq:new_optimization_start}-\ref{eq:new_optimization_end}) non-analytic via theoretical derivations.
Please refer to Supplementary Figure 11 for an illustration on system resilience of a cellular network.

\Methodsubsection{Neuronal dynamics}\label{method::neuronal}

We then investigate the neuronal dynamics describing the interactions among neurons.
The activity of neurons are characterized by the following Wilson–Cowan dynamics\cite{wilson1972excitatory,wilson1973mathematical},
\begin{equation}
    \frac{\mathrm{d}x_i}{\mathrm{d}t} = -b_i x_i + \sum_{j=1}^N{A_{ij}\frac{1}{1+e^{\mu - \delta x_j}}},\label{eq:method_neuronal}
\end{equation}
where self-dynamics $F_i = -b_i x_i$ and interactions $G_{ij} = \frac{1}{1+e^{\mu - \delta x_j}}$ also display strong heterogeneity and non-linearity.
The dynamical heterogeneity is achieved similarly by a power-law on $b_i$ as equation (\ref{eq:power_law}).

In contrast to cellular dynamics, a neuronal network can feature bifurcation and inactivity under external perturbations, which both indicate that the network loses its resilience\cite{gao2016universal,sanhedrai2022reviving}.
Therefore, to comprehensively measure network resilience, we need to compute the steady states involving environmental changes.
Specifically, we can compare the steady states obtained from different initial states, where the distinction represents perturbations.
Here we include two initial states, low and high, and their corresponding steady states are computed as follows,
\begin{align}
    &\mathbf{x}_{0}^H = (x_1, \ldots, x_N)_{t=0}^H = (10, \ldots, 10), \quad \mathbf{x}_{0}^L = (x_1, \ldots, x_N)_{t=0}^L = (0, \ldots, 0)\\
    &\mathbf{x}_{T}^H = \mathbf{x}_{0}^H + \int_0^T -\mathbf{b}\mathbf{x} + \mathbf{A}\frac{1}{1 + e^{\mu - \delta \mathbf{x}}}\mathrm{d}\mathbf{x}, \quad \mathbf{x}_{T}^L = \mathbf{x}_{0}^L + \int_0^T -\mathbf{b}\mathbf{x} + \mathbf{A}\frac{1}{1 + e^{\mu - \delta \mathbf{x}}}\mathrm{d}\mathbf{x},\label{eq:neuron_integration}
\end{align}
where $\mathbf{b}=(b_1, \ldots, b_N)$ is the heterogeneous dynamic parameters and $T=400$ to obtain the steady state.
A neuronal network is non-resilient either the two steady states feature bifurcation or the system becomes inactive, otherwise it is resilient under perturbations.
Thus the resilience function is defined by the following equation,
\begin{align}
    \mathcal{R}(\mathbf{A}, [\mathbf{F}, \mathbf{G}]) = \begin{cases} 1 & \text{if } \mathbf{x}_T^H = \mathbf{x}_T^L \text{ and } \langle \mathbf{x}_T^L \rangle > 0 \\ 0 & \text{if }  \mathbf{x}_T^H \neq \mathbf{x}_T^L \text{  or  } \langle \mathbf{x}_T^L \rangle = 0 \end{cases}\label{eq:resilience_neuron}
\end{align}
Similarly, the resilience function (\ref{eq:resilience_neuron}) is also highly non-analytic, which can only be treated in a computational rather than theoretical way.
Please refer to Supplementary Figure 12 for an illustration on system resilience of a cellular network.

\bibliography{references}

\section*{Author contributions}

Y.Z., J.D., and Y.L. conceived of the project and designed the research methods.
Y.Z. developed the self-inductive symbolized reinforcement learning framework and performed the experiments, with J.D., D.J., J.G., and Y.L. contributing technical advice and ideas. 
J.D. and Y.L. directed and managed the project.
All authors jointly analyzed the results and participated in the writing of the manuscript.

\section*{Competing interests}
The authors declare no competing interests.

\section*{Code availability}

The code used in this research has been made available on GitHub (\url{https://github.com/tsinghua-fib-lab/selinda}).

\section*{Data availability}

The datasets used in the experiments have been made available on GitHub (\url{https://github.com/tsinghua-fib-lab/selinda}).

\newpage

\newpage
\begin{table}[ht]
\centering
\caption{\textbf{Evaluation of the discovered theoretical formulas of network resilience by our approach compared with existing methods.}
We report the removal cost on real-world large-scale networks with cellular and neuronal dynamics across different types and sizes.}\label{tab:result}
\footnotesize
\begin{tabular}{c|cc|cccccc|cc|cc}
\toprule
\multirow{2}{*}{\bf{Dynamic}} & \multirow{2}{*}{\bf{Network Type}} & \multirow{2}{*}{\bf{\textit{N}}} & \multicolumn{6}{c|}{\bf{Theoretical Method}} & \multicolumn{2}{c|}{\bf{AI Method}} & \multirow{2}{*}{\bf{Ours}} & \multirow{2}{*}{\bf{impr\%}} \\
& & & \bf{DC\cite{albert2000error}} & \bf{RC\cite{zhang2020resilience}} & \bf{CI\cite{morone2015influence}} & \bf{GND\cite{ren2019generalized}} & \bf{EI\cite{clusella2016immunization}} & \bf{CoreHD\cite{zdeborova2016fast}} & \bf{GDM\cite{grassia2021machine}} & \bf{FINDER\cite{fan2020finding}} & & \\
\midrule
\multirow{8}{*}{\bf{Cellular}} & \multirow{4}{*}{Human} & 355 & 9 & 9 & 14 & 174 & 78 & 37 & 34 & 9 & \bf{4} & \bf{55.6\%} \\
& & 549 & \underline{22} & 23 & 23 & 259 & 171 & 100 & 82 & \underline{22} & \bf{11} & \bf{50.0\%} \\
& & 978 & 64 & 64 & 71 & 557 & 470 & 248 & 195 & 67 & \bf{32} & \bf{50.0\%} \\
& & 3125 & 272 & \underline{256} & 293 & 2026 & 1114 & 726 & 507 & 263 & \bf{125} & \bf{51.2\%} \\
\cline{2-13}
& \multirow{4}{*}{Yeast} & 491 & 17 & 16 & 18 & 213 & 176 & 101 & 24 & 17 & \bf{11} & \bf{31.3\%} \\
& & 701 & 30 & 38 & 32 & 281 & 240 & 141 & 51 & 31 & \bf{8} & \bf{73.3\%} \\
& & 931 & 50 & 43 & 61 & 403 & 343 & 209 & 80 & 52 & \bf{16} & \bf{62.8\%} \\
& & 1647 & 121 & 118 & 116 & 808 & 661 & 398 & 193 & 117 & \bf{30} & \bf{74.1\%} \\
\hline
\multirow{4}{*}{\bf{Neuronal}} & \multirow{4}{*}{Brain} & 286 & 103 & 105 & 200 & 180 & 258 & 232 & 203 & 205 & \bf{84} & \bf{18.4\%} \\
& & 359 & 158 & 155 & 209 & 236 & 267 & 302 & 251 & 210 & \bf{122} & \bf{21.3\%} \\
& & 676 & 278 & \underline{271} & 476 & 351 & 541 & 570 & 490 & 446 & \bf{222} & \bf{18.1\%} \\
& & 989 & \underline{428} & 430 & 667 & 490 & 889 & 854 & 745 & 543 & \bf{335} & \bf{21.7\%} \\
\bottomrule
\end{tabular}
\end{table}

\newpage
\begin{figure}[ht]
\centering
\includegraphics[width=\linewidth]{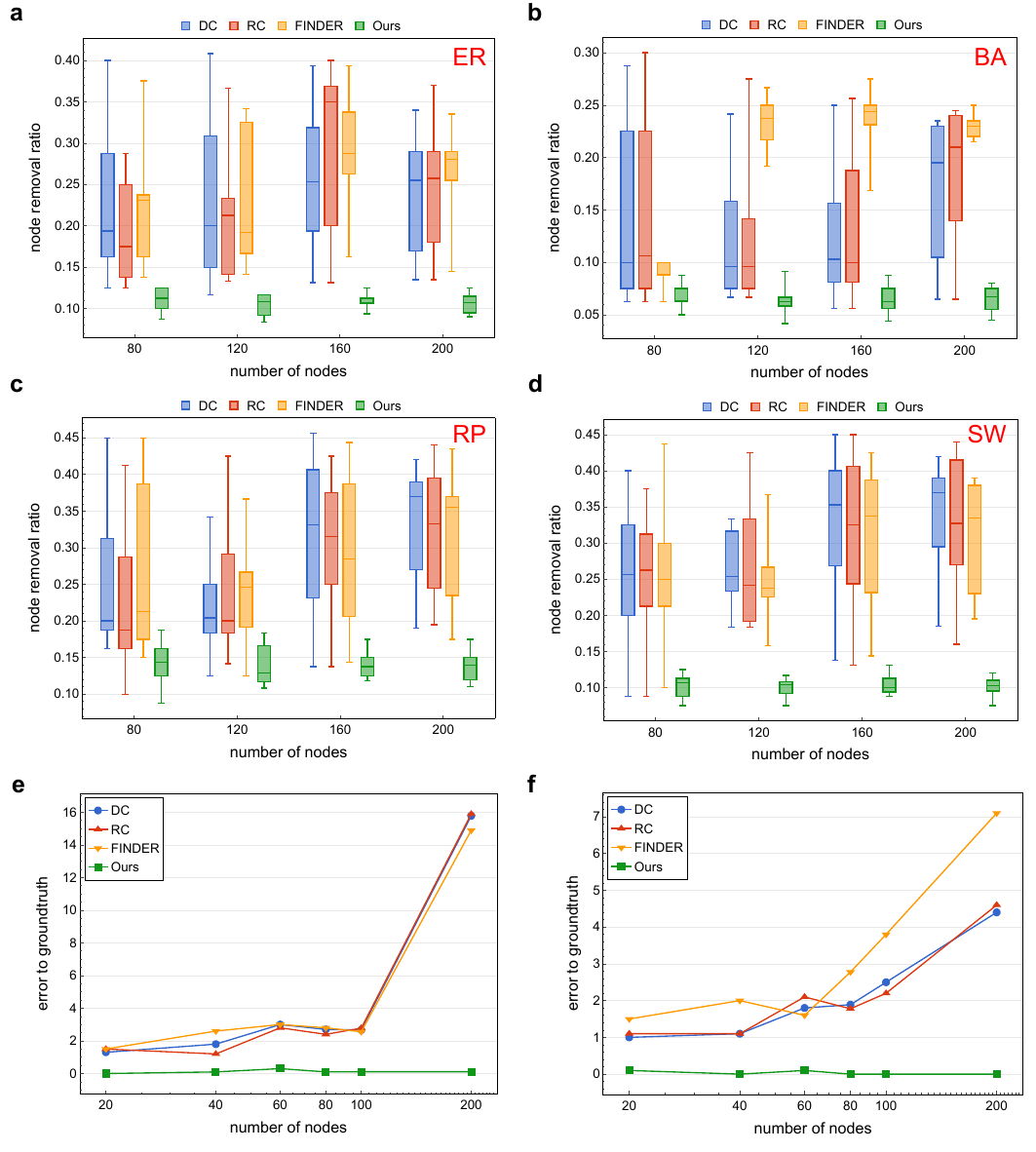}
\caption{\textbf{a-d.} The node removal ratio of different methods (lower means more accurate) for \textbf{a.} ER networks \textbf{b.} BA networks \textbf{c.} RP networks and \textbf{d.} SW networks of growing network sizes, with each size containing 10 synthetic networks driven by gene regulatory dynamics.
Results demonstrate the robustness of our discovered theory, which constantly provides precise identification of keystone nodes, with the number of removed nodes reduced by over 43.4\% in comparison to existing approaches.
\textbf{e-f.} The average error between the node removal ratio and the ground-truth value of different methods for \textbf{e.} cellular dynamics and \textbf{f.} neuronal dynamics of growing network sizes, with each size containing 10 synthetic ER networks.
Results illustrate that existing methods exhibit significant errors as network size increases, with the largest overestimation of keystone nodes by over 7.5\%.
In contrast, our discovered theory constantly delivers optimal accuracy with near-zero errors to the groundtruth values in removal cost regardless of network size.
}
\label{fig:result1}
\end{figure}

\newpage
\begin{figure}[ht]
\centering
\includegraphics[width=0.95\linewidth]{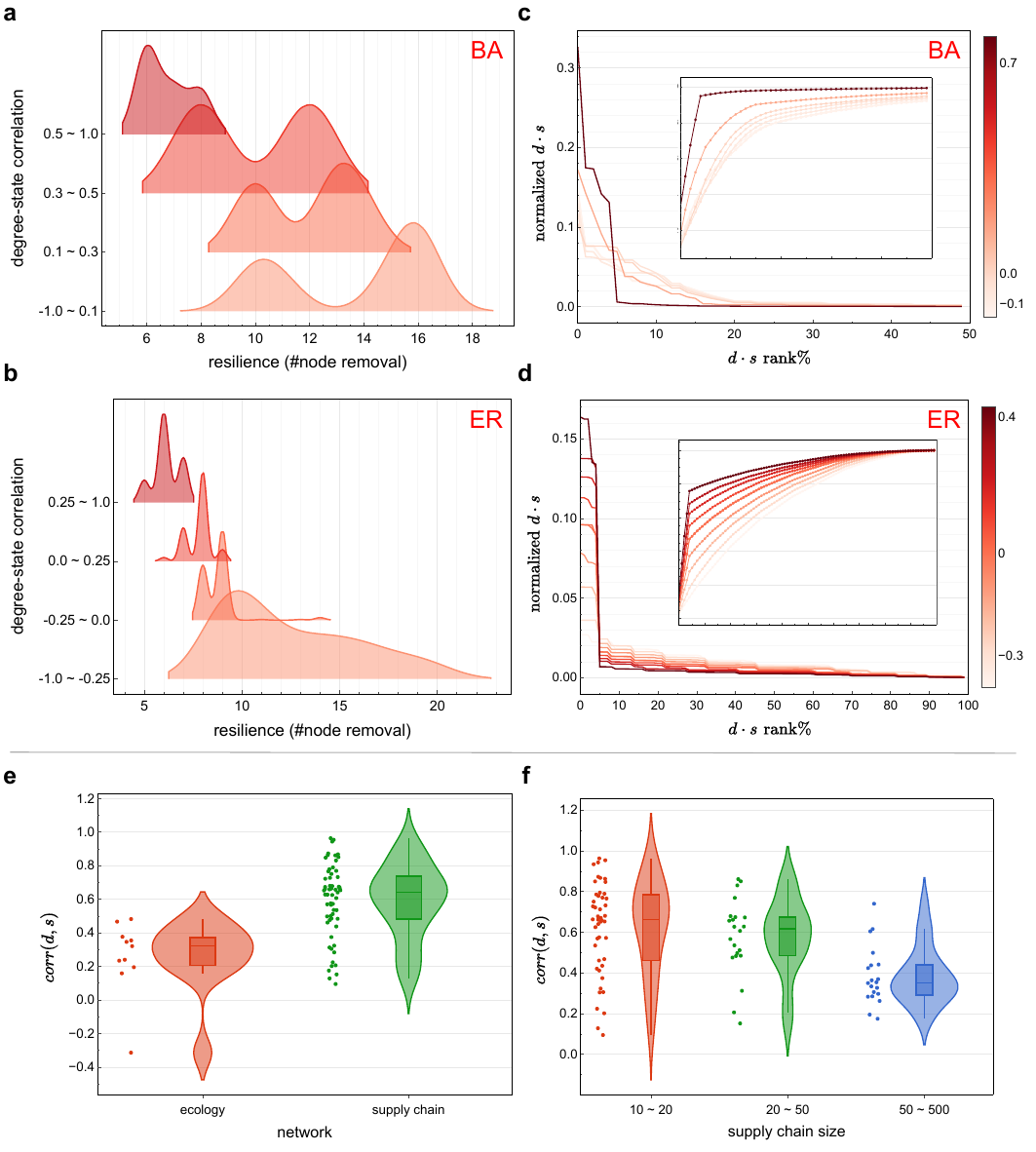}
\vspace{-15px}
\caption{
\textbf{a-b.} Resilience measured by the number of removed nodes under different correlations between node degree distribution and node state distribution for \textbf{a.} neuronal BA networks and \textbf{b.} neuronal ER networks. The dynamical parameters are set to different values to achieve different degree-state correlation. 
The results demonstrate negative relation between degree-state correlation and network resilience, where networks are least resilient under high $\texttt{corr}(d, s)$ indicating correlated distribution of degree and state, and decreasing $\texttt{corr}(d, s)$ can lead to resilience improvements of over 183\% and 300\%.
\textbf{c-d.} The mass of $d \cdot s$ over different nodes under different degree-state correlation for \textbf{c.} neuronal BA networks and \textbf{d.} neuronal ER networks. We display the cumulative distribution of $d \cdot s$ in the picture in picture.
The mass of $d \cdot s$, which indicates the risk of network collapse, is not concentrated on hub nodes but rather distributed more uniformly across a wide range of nodes under low $\texttt{corr}(d, s)$, thereby requiring more nodes to be removed to induce the loss of resilience.
\textbf{e.} $\texttt{corr}(d, s)$ of real-world ecology networks and supply chain networks. $s$ indicates specie abundance and company throughput for the two categories of networks, respectively.
The $\texttt{corr}(d, s)$ difference between ecology and supply chain networks is over 55.9\%, which is consistent with common beliefs that natural systems are more resilience than artificial ones.
\textbf{f.} $\texttt{corr}(d, s)$ of real-world supply chain networks of different sizes.
Increasing the size of supply chain networks can lead to a 38.8\% drop in $\texttt{corr}(d, s)$, suggesting potential resilience benefits from incorporating diverse additional suppliers.
}
\label{fig:result2}
\end{figure}

\clearpage
\newpage
\begin{figure}[ht]
\centering
\includegraphics[width=\linewidth]{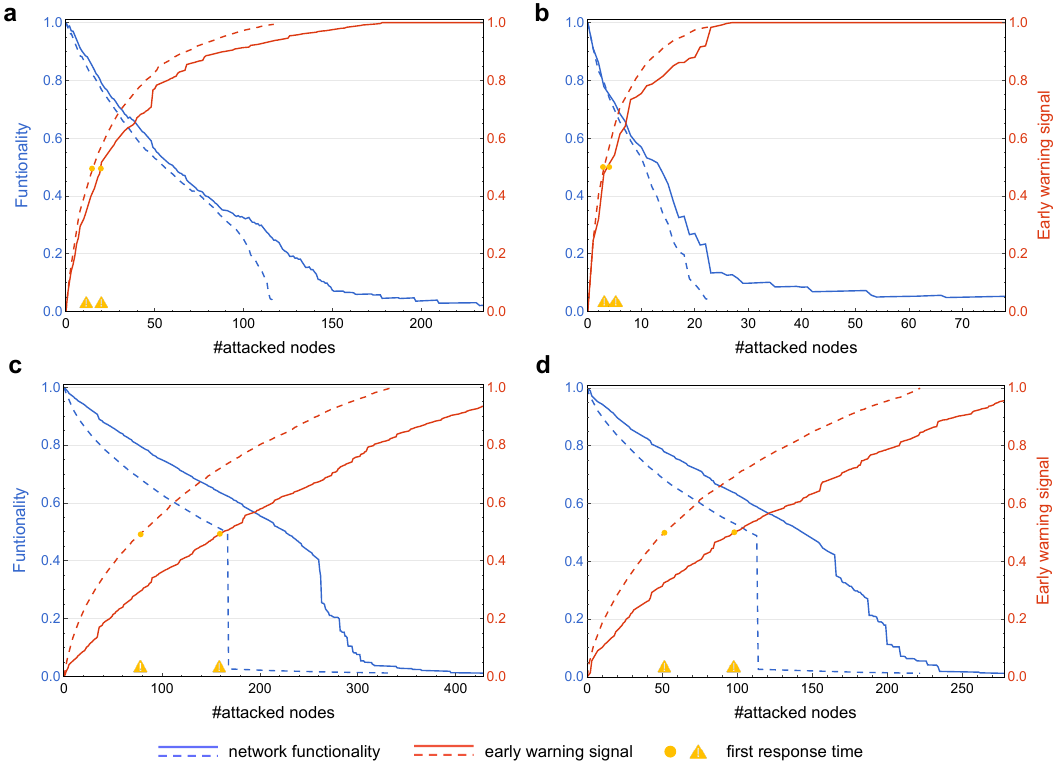}
\caption{
Network functionality (blue) and early warning signal $\alpha$ (red) under DC attack strategy (solid) that removes hub nodes first and $d \cdot s$ attack strategy (dashed) that prioritizes nodes with larger $d \cdot s$.
The first response time (orange circle and exclamation mark) indicates the occurrence of 50\% early warning.
Experiments are conducted on \textbf{a.} A human gene regulatory network,
\textbf{b.} A yeast gene regulatory network, and \textbf{c-d.} Two brain neuronal networks.
The results show that the first response time arises long before the network disruption with the relative time 38\% ahead of the actual collapse, thus it allows sufficient time for emergency responses, proving the effectiveness of our proposed early warning signals in preventing catastrophic collapses.
}
\label{fig:result3}
\end{figure}

\begin{table}[ht]
\begin{minipage}[t]{0.45\textwidth}
\footnotesize
\begin{tabular}{l|ccc}
\toprule
\bf{Method} & \makecell[c]{\bf{Crime} \\ \bf{(829)}} & \makecell[c]{\bf{HI-II-14} \\ \bf{(4165)}} & \makecell[c]{\bf{Facebook} \\ \bf{(63392)}} \\
\midrule
DC\cite{albert2000error}: $d$ & 0.1242 & 0.0691 & 0.3165 \\
CI\cite{morone2015influence}: $(k_i - 1) \sum_{j \in \partial \text{Ball}(i, l)} (k_j - 1)$ & \underline{0.1132} & \underline{0.0561} & \underline{0.2339} \\
GND\cite{ren2019generalized} & 0.2570 & 0.3875 & 0.4311 \\
EI\cite{clusella2016immunization}: $k_i^{(\text{eff})} + \sum_{\mathcal{C} \subset \mathcal{N}_i} \left( \sqrt{|\mathcal{C}|} - 1 \right)$ & 0.3259 & 0.1400 & 0.3632 \\
CoreHD\cite{zdeborova2016fast} & 0.2061 & 0.1155 & 0.3174 \\
GDM\cite{grassia2021machine} & 0.1168 & 0.0749 & 0.2994 \\
FINDER\cite{fan2020finding} & 0.1210 & 0.0645 & 0.3095 \\
\midrule
Ours: $d^2 / \bar{d}$ & \bf{0.1092} & \bf{0.0559} & \bf{0.2309} \\
\bottomrule
\end{tabular}
\caption{Evaluation of the discovered theoretical formulas of network resilience by our approach compared with existing methods of the past 10 years.
We report the removal cost on real-world large-scale social networks, where our discovered theoretical formula achieves the best performance in identifying keystone nodes across 3 networks of different sizes, with a maximum of 3.53\% improvements.}
\label{tab:null-dynamics}
\end{minipage}
\hspace{0.02\textwidth}
\begin{minipage}[t]{0.45\linewidth}
\centering
\small
\begin{tabular}{c|c|c|c|c}
\toprule
\bf{Dynamic} & \bf{Network} & \bf{\textit{N}} & \makecell[c]{\bf{RC\cite{zhang2020resilience}:} \\ \bf{$2\bar{d} + d(d - 2\beta)$}} & \makecell[c]{\bf{Ours:} \\ \bf{$2\bar{d} + d(\bar{d} - 2\beta)$}} \\
\midrule
\multirow{6}{*}{Cellular} & \multirow{3}{*}{Human} & 355 & 8 & \bf{5} \\
& & 978 & 45 & \bf{35} \\
& & 3125 & 153 & \bf{141} \\
\cline{2-5}
& \multirow{3}{*}{Yeast} & 491 & 16 & \bf{13} \\
& & 931 & 33 & \bf{28} \\
& & 1647 & 54 & \bf{53} \\
\hline
\multirow{2}{*}{Neuronal} & \multirow{2}{*}{Brain} & 286 & 104 & \bf{103} \\
& & 676 & 269 & \bf{268} \\
\bottomrule
\end{tabular}
\caption{Evaluation of the formula discovered by our method in comparison to the state-of-the-art RC theory on real-world large-scale networks with homogeneous dynamics.
Our discovered formula improves over RC by 22.5\% and 11.9\% in accuracy of keystone node identification for both human and yeast networks driven by cellular dynamics with a maximum improvement of 37.5\%, and outperforms RC on two brain networks driven by neuronal dynamics.}
\label{tab:homo-dynamics}
\end{minipage}
\end{table}

\end{document}